\documentstyle[jkas]{article}

\beginpage{1}
\endpage{8}
\year{2013}\volume{00}\month{May}

\runningauthor {S. Trippe} 
\runningtitle{Can Massive Gravity Explain the MDA Relation?} 

\month{May} \year{2013} \volume{00} \issueno{00}
\beginpage{1}\endpage{8}
\date{Received 2013 April 15; Revised 2013 May 27; Accepted 2013 May 27}

\begin{document}

\def\Mo{{M_{0}}}
\def\Me{{M_{\rm ex}}}
\def\Mt{{M_{\rm tot}}}
\def\ax{{a_{\rm x}}}
\def\ao{{a_0}}
\def\ac{{a_{\rm c}}}
\def\acp{{a'_{\rm c}}}
\def\am{{a_{\rm M}}}
\def\gn{{g_{\rm N}}}
\def\vc{{v_{\rm c}}}
\def\vo{{v_{0}}}
\def\mg{{m_{\rm g}}}
\def\al{{\langle}}
\def\ar{{\rangle}}
\def\deg{{^{\circ}}}
\def\xii{{\xi_{1}(x)}}
\def\xiii{{\xi_{2}(x)}}
\def\xiiii{{\xi_{3}(x)}}
\def\xin{{\xi_{n}(x)}}
\def\xie{{\xi_{\rm e}(x)}}

\title{CAN MASSIVE GRAVITY EXPLAIN THE MASS DISCREPANCY--ACCELERATION RELATION OF DISK GALAXIES?}

\author{Sascha Trippe}

\address{Seoul National University, Department of Physics and Astronomy, Seoul 151-742, Korea\\ {\it E-mail : trippe@astro.snu.ac.kr}}

\address{\normalsize{\it (Received 2013 April 15; Revised 2013 May 27; Accepted 2013 May 27)}}

\abstract{
\noindent The empirical mass discrepancy--acceleration (MDA) relation of disk galaxies provides a key test for models of galactic dynamics. In terms of modified laws of gravity and/or inertia, the MDA relation quantifies the transition from Newtonian to modified dynamics at low centripetal accelerations $a_c\lesssim10^{-10}$\,m\,s$^{-2}$. As yet, neither dynamical models based on dark matter nor proposed modifications of the laws of gravity/inertia have predicted the functional form of the MDA relation. In this work, I revisit the MDA data and compare them to four different theoretical scaling laws. Three of these scaling laws are entirely empirical, the fourth one -- the ``simple $\mu$'' function of Modified Newtonian Dynamics -- derives from a toy model of gravity based on massive gravitons (the ``graviton picture''). All theoretical MDA relations comprise one free parameter of the dimension of an acceleration, Milgrom's constant $\am$. I find that the ``simple $\mu$'' function provides a good fit to the data free of notable systematic residuals and provides the best fit among the four scaling laws tested. The best-fit value of Milgrom's constant is $\am=(1.06\pm0.05)\times10^{-10}$\,m\,s$^{-2}$. Given the successful prediction of the functional form of the MDA relation, plus an overall agreement with the observed kinematics of stellar systems spanning eight orders of magnitude in size and 14 orders of magnitude in mass, I conclude that the ``graviton picture'' is sufficient (albeit probably not a necessary or unique approach) to describe galactic dynamics on all scales well beyond the scale of the solar system. This suggests that, at least on galactic scales, gravity behaves as if it was mediated by massive particles.
}


\maketitle

\section{INTRODUCTION \label{sect_intro}}

\noindent
Since the seminal works by \citet{zwicky1933}, \citet{gallagher1976}, and \citet{rubin1980} it has become evident (e.g., \citealt{binney1987,sanders1990}) that the dynamical masses of galaxies and galaxy clusters exceed their luminous masses by up to one order of magnitude -- the well-known \emph{missing mass problem}. On the one hand, this observation led to the postulate of non-luminous and non-baryonic \emph{dark matter} \citep{ostriker1973,einasto1974}, eventually evolving into the $\Lambda$CDM standard model of cosmology (e.g., \citealt{bahcall1999}). On the other hand, various proposals for \emph{modified laws of gravity and/or inertia} have been made to address the missing mass problem; the most successful candidate to date appears to be \emph{Modified Newtonian Dynamics} (MOND; \citealt{milgrom1983a,milgrom1983b,milgrom1983c,bekenstein1984}; also \citealt{bekenstein2006,ferreira2009}). As yet, MOND provides the best description of galactic dynamics (see \citealt{famaey2012} for a recent review; cf. also \citealt{kroupa2012}). MOND postulates a modification of the laws of gravity and/or inertia such that the expression for the circular speed $\vc$ of a test mass orbiting a luminous mass $\Mo$ transits from the usual Newtonian form $\vc^2=G\Mo/r$ to $\vc^4=G\,\Mo\,\am=const.$ when approaching very low centripetal accelerations $\ac\ll10^{-10}$\,m\,s$^{-2}$; here $r$ is the radial distance from $\Mo$, $G$ is Newton's constant and $\am\approx10^{-10}$\,m\,s$^{-2}$ is \emph{Milgrom's constant}.

Even though MOND provides the correct limiting cases by construction, it does not provide the functional form of the transition from Newtonian to modified dynamics. Nevertheless, this transitional regime has been explored observationally, leading to the discovery of the empirical \emph{mass discrepancy--acceleration (MDA) relation} of disk galaxies \citep{mcgaugh2004}. The MDA relation comprises a universal scaling of the ratio $\Mt/\Mo$ -- with $\Mt$ and $\Mo$ being the dynamical and baryonic masses respectively -- with centripetal acceleration $\ac$. Substantial deviations of $\Mt/\Mo$ from unity occur at $\ac\lesssim10^{-10}$\,m\,s$^{-2}$. As yet, neither dynamical models based on dark matter nor proposed laws of modified gravity/inertia have predicted the functional form of the MDA relation. Even though the functional form of the MDA relation is not provided by theory, it is constrained by several boundary conditions. First, any theoretical relation must comprise (a) Newtonian dynamics at high accelerations and (b) MONDian dynamics at low accelerations. Second, the theoretical MDA relation has to connect the two limiting cases unambiguously and smoothly.

In this paper, I revisit the empirical MDA relation and compare it to four different theoretical scaling laws known from the description of galactic rotation curves. One of these scaling laws can be derived from a toy model of gravity, the ``graviton picture'', based on massive gravitons; the other three are entirely empirical. Based on a comparison of fit residuals, I argue that the scaling law derived from the ``graviton picture'' provides the best fit to the data. This suggests that, at least on galactic scales, gravity behaves as if it was mediated by massive particles.

\section{ANALYSIS}

\subsection{Observational MDA Data}

\noindent
I make use of the MDA data set of \citet{sanders2002}, \citet{mcgaugh2004}, and \citet{famaey2012} as summarized in Fig.~10 of \citet{famaey2012}.\footnote{Made publicly available by {\sc Stacy S. McGaugh} via {\tt http://astroweb.case.edu/ssm/data/MDaccRgn\textunderscore LR.dat} .} For each galaxy, total baryonic mass (stars and gas) and circular velocity are derived as function of radial distance from the galactic center, $r$. Velocities are determined from optical and/or radio line spectroscopy of hydrogen, baryonic masses from optical/near-infrared photometry and H\,{\sc i} 21-cm radio line mapping.  

For a given radial distance $r$, the mass discrepancy is derived from comparison of observed circular speed, $\vc$, with the speed expected from the enclosed baryonic mass, $\vo= \sqrt{G\Mo/r}$; $G$ is Newton's constant. The \emph{mass discrepancy} (MD) is then given by

\begin{equation}
\label{eq_masses}
\frac{\Mt}{\Mo} = \left(\frac{\vc}{\vo}\right)^2 ~ .
\end{equation}

\noindent
This ratio can be analyzed either as function of the centripetal acceleration $\ac=\vc^2/r$ (``representation 1'', ``R1'') or as function of the Newtonian acceleration expected for the case $\Mt=\Mo$, i.e., $\gn=\vo^2/r$ (``representation 2'', ``R2''). Empirically, the MD is found to anti-correlate with $\ac$ and $\gn$ but to not correlate with distance $r$ or orbital frequency \citep{mcgaugh2004}. The empirical MDA relation is illustrated in Fig.~10 of \citet{famaey2012} which comprises 735 measurements from a sample of 60 galaxies. For all data points, the relative statistical precisions are better than 5\% for $\vc$ and thus better than 7\% for $\Mt\propto\vc^2$.

\subsection{Theoretical MDA Relation \label{ssect_theomda}}

\noindent
According to the discussion provided in \citet{trippe2013a,trippe2013b}, it is possible to derive a theoretical MDA relation from a toy model of gravity. This model, the ``graviton picture'', assumes \emph{ad hoc} that (1) gravitation is mediated by discrete particles, \emph{gravitons}; (2) these gravitons are virtual particles arising from quantum fluctuations, meaning they do not remove energy from the emitting mass; (3) gravitons have a non-zero mass; and (4) graviton--graviton interactions are excluded. The ``graviton picture'' is an unusual though -- a priori -- straightforward extrapolation of the standard assumptions made in quantum field theories (cf., e.g., \citealt{griffith2008} for an overview; see also \citealt{goldhaber2010,hinterbichler2012} for reviews on massive gravitons). From assumptions 1--4 it follows that a baryonic source mass $\Mo$ radiates away massive gravitons, thus forming a (electromagnetically dark) graviton halo with mass density profile

\begin{equation}
\label{eq_halo}
\rho(R) = \Mo\,\beta\,R^{-2}
\end{equation}

\noindent
where $R$ denotes the radial distance from $\Mo$ and $\beta$ is a scaling parameter of the dimension of an inverse length. The proportionality $\rho\propto\Mo$ follows from consistency with classical field theory; the proportionality $\rho\propto R^{-2}$ follows from the inverse-square-of-distance law of radiation. Assuming a test particle orbiting $\Mo$ on a circular orbit with radius $r$ and circular speed $\vc$, one can re-write $\beta$ like

\begin{equation}
\label{eq_beta}
\beta = \frac{2\ao}{\vc^2}
\end{equation}

\noindent
with $\vc^2/2$ being the kinetic energy per unit mass of the test particle and $\ao$ denoting a constant of the dimension of an acceleration. When integrating $\rho(R)$ over $R$ from 0 to $r$ in spherical coordinates and adding $\Mo$, one finds the theoretical MDA relation

\begin{equation}
\label{eq_mda}
\frac{\Mt}{\Mo} = 1 + \frac{\am}{\ac} \equiv \xii = 1 + \frac{1}{x}
\end{equation}

\noindent
with $\am=8\pi\ao$ being Milgrom's constant and $x=\ac/\am$ \citep{trippe2013a,trippe2013b}. Milgrom's constant is the only free parameter of the model; $\Mt/\Mo$ remains very close to unity as long as $\ac\gg\am$. Notably, the empirical MDA relation implies that $\am$ is a \emph{universal} constant of nature which is identical for \emph{all} galactic systems (cf. \citealt{famaey2012}). The Newtonian acceleration $\gn$ is given by $\gn=\ac/(\Mt/\Mo)$  by construction. Expressing the theoretical MD as function of either $\ac$ (as in Eq.~\ref{eq_mda}) or $\gn$ leads to the empirical ``simple $\mu$'' and ``simple $\nu$'' functions commonly employed in MOND, respectively \citep{famaey2012,trippe2013b}.

\subsection{Alternative MDA Scaling Laws}

\noindent
In addition to $\xii$, several other scaling laws can be, and have been, applied to galactic rotation curves (e.g., \citealt{milgrom1983b,famaey2005,famaey2012}). As noted by, e.g., \citet{famaey2012}, the boundary conditions discussed in \S\,\ref{sect_intro} are fulfilled by the family of functions

\begin{equation}
\label{eq_xin}
\xin = \left[1 + \left(\frac{1}{x}\right)^n\right]^{1/n} ~ , ~~ n = 1, 2, 3, ... ~ ,
\end{equation}

\noindent
with $x=\ac/\am$ as before. The cases $n=1, 2$ have been used to successfully model galactic rotation curves (e.g., \citealt{famaey2005}). As already noted by \citet{milgrom1983b}, an additional valid MDA scaling law is given by

\begin{equation}
\label{eq_xiexp}
\xie = \frac{1}{1 - {\rm e}^{-x}} ~~ , ~~~ {\rm e} = 2.718... ~~ .
\end{equation}

In the following discussion, I consider the scaling laws $\xii$, $\xiii$, $\xiiii$, and $\xie$. Else than $\xii$, which I discuss in \S\,\ref{ssect_theomda}, the other three scaling laws are entirely empirical and do not derive from physical models; technically, it is possible to construct arbitrary additional scaling laws from the known boundary conditions.

\subsection{Model vs. Data \label{ssect_comparison}}

\noindent
Having the empirical as well as a theoretical MDA relation at hand, comparison between data and model is, a priori, straightforward. However, one encounters some complications affecting the observational MDA data: First, any empirical value for $\Mt/\Mo$ depends on the choice of a mass-to-light ratio made for a specific galaxy. This choice is inevitably affected by a systematic uncertainty which introduces a systematic scatter into the empirical MDA relation that partially dominates over the formal statistical errors. Second, data at high accelerations $\gn\gtrsim10^{-10}$\,m\,s$^{-2}$ are obtained from high surface brightness galaxies at low radii where the velocities gradients ${\rm d}\vc/{\rm d}r$ are steep and accurate measurements of $\vc$ are difficult. Third, the empirical MDA relation assumes circular motion which is a good though non-perfect approximation \citep{mcgaugh2004}. Careful inspection of the data (cf. Fig.~10 of \citealt{famaey2012}) shows that the high-acceleration tails of the MDA distributions drop slightly below unity -- indicating a bias in the data which one needs to take into account. Eventually, I probe the agreement of data and model according to the following scheme:

\begin{enumerate}

\item  Make a first guess for $\am$ such that the model curves follow the data approximately.

\item  Add a constant offset $\delta$ to $\Mt/\Mo$ such that (a) the trend lines of the empirical distributions are placed above unity for all accelerations and (b) empirical distributions and theoretical curves connect smoothly at high accelerations  $\ac\approx\gn\gtrsim3\times10^{-10}$\,m\,s$^{-2}$. As $\ac\propto\Mt$ and $\gn\propto\Mo$, actual centripetal acceleration $\ac$ and measured acceleration $\acp$ are related like $\acp=\ac-\delta\times\gn$.

\item  Optimize the choice for $\am$ such that the root-mean-squared (r.m.s.) residual difference between model and data, $\eta$, is minimized in R2.

\end{enumerate}

Steps 1 and 2 are applied to R1 and R2 simultaneously, step 3 is applied to R2 only because the scatter is lower in R2 than in R1.

\section{RESULTS \label{sect_results}}

\begin{figure*}[t!]
\centering
\includegraphics[angle=-90,width=155mm]{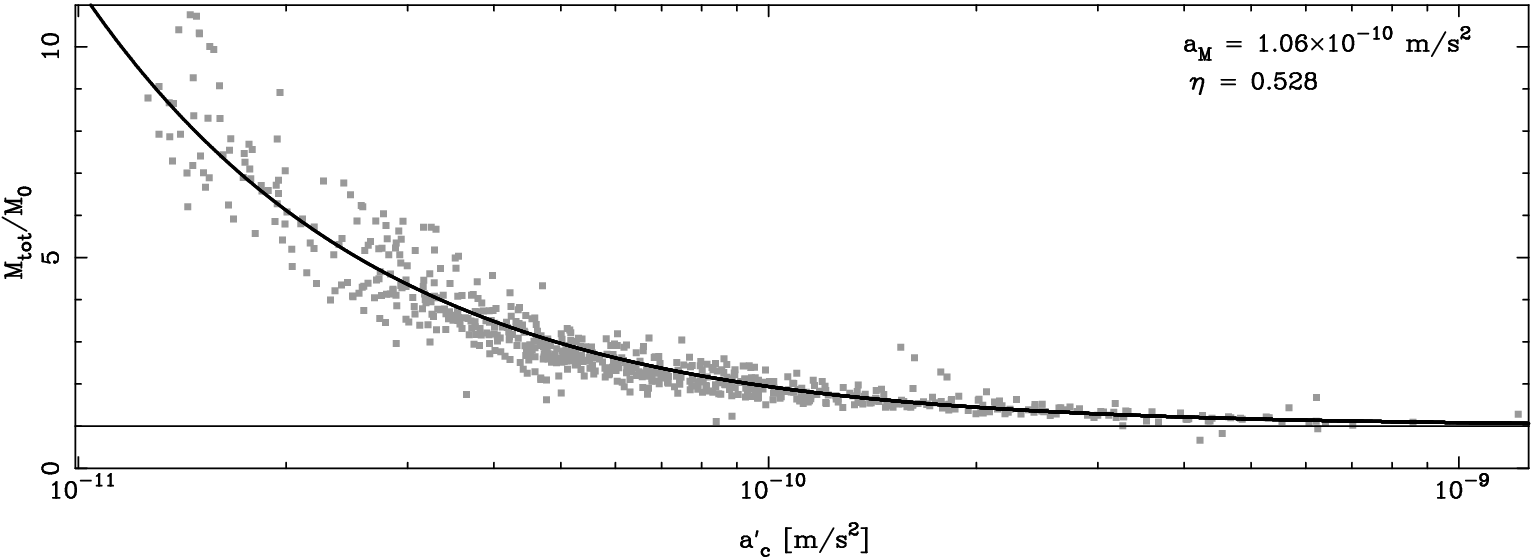} \\
\vspace{5mm}
\includegraphics[angle=-90,width=155mm]{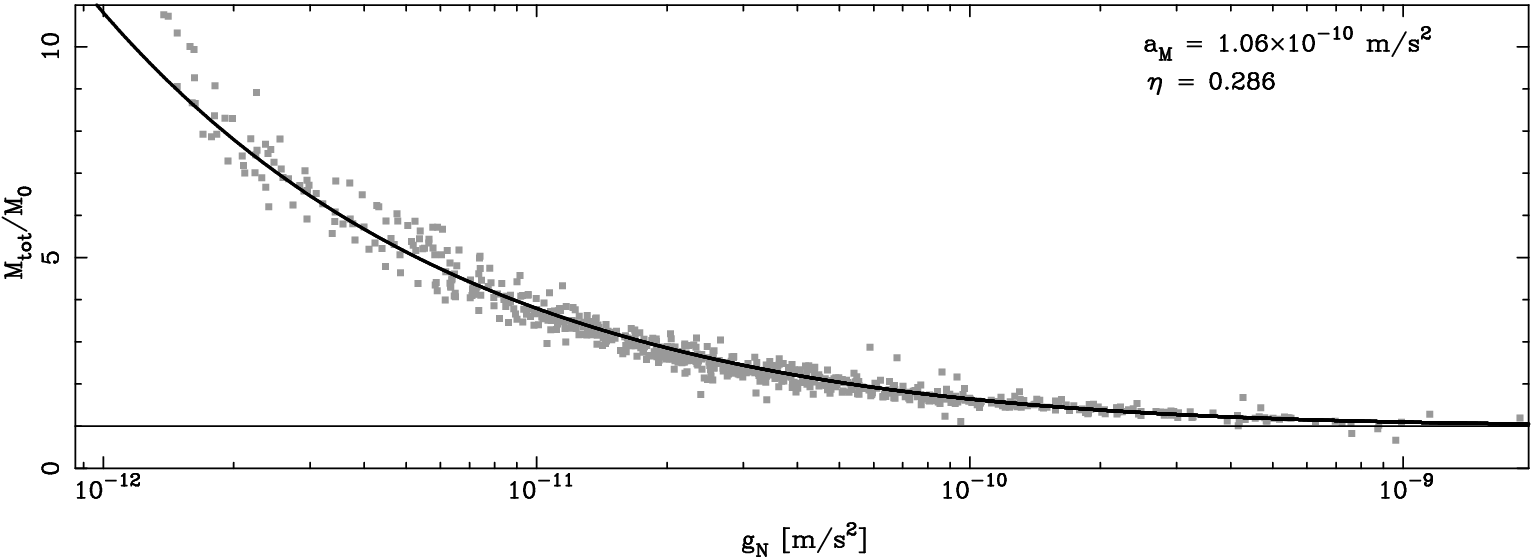}
\vspace{2mm}
\caption{Comparison of theoretical and empirical MDA relations. Here $\Mt$ is the total dynamical mass, $\Mo$ is the luminous baryonic mass. The mass discrepancy is given by the ratio $\Mt/\Mo$. \emph{Top panel}: Mass discrepancy vs. measured centripetal acceleration $\acp$. \emph{Bottom panel:} Mass discrepancy vs. Newtonian acceleration $\gn$, i.e., the acceleration expected if $\Mt=\Mo$. Grey squares indicate the bias-corrected ($\delta=0.225$) MDA data by \citet{famaey2012}, the continuous black curve corresponds to $\xii$ (Eq.~\ref{eq_mda}) for a characteristic acceleration $\am=1.06\times10^{-10}$\,m\,s$^{-2}$. Please note the logarithmic--linear axis scales. Data and model are in good agreement, the r.m.s. residuals are $\eta=0.53$ and $\eta=0.29$ (in units of $\Mt/\Mo$), respectively. \label{fig_mda}}
\end{figure*}

\begin{figure*}[t!]
\centering
\includegraphics[angle=-90,trim=0mm 0mm 15mm 0mm,clip,width=80mm]{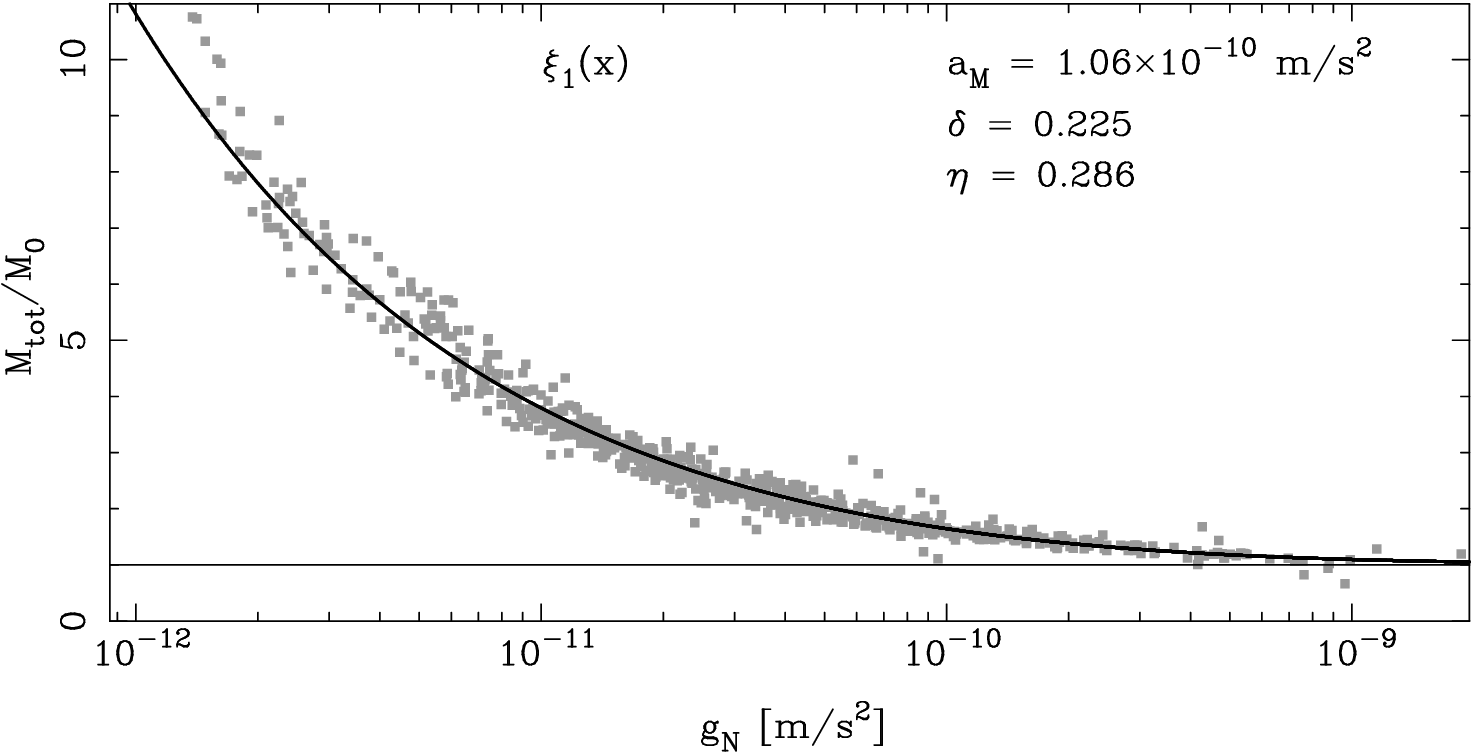}
\hspace{5mm}
\includegraphics[angle=-90,trim=0mm 0mm 15mm 0mm,clip,width=80mm]{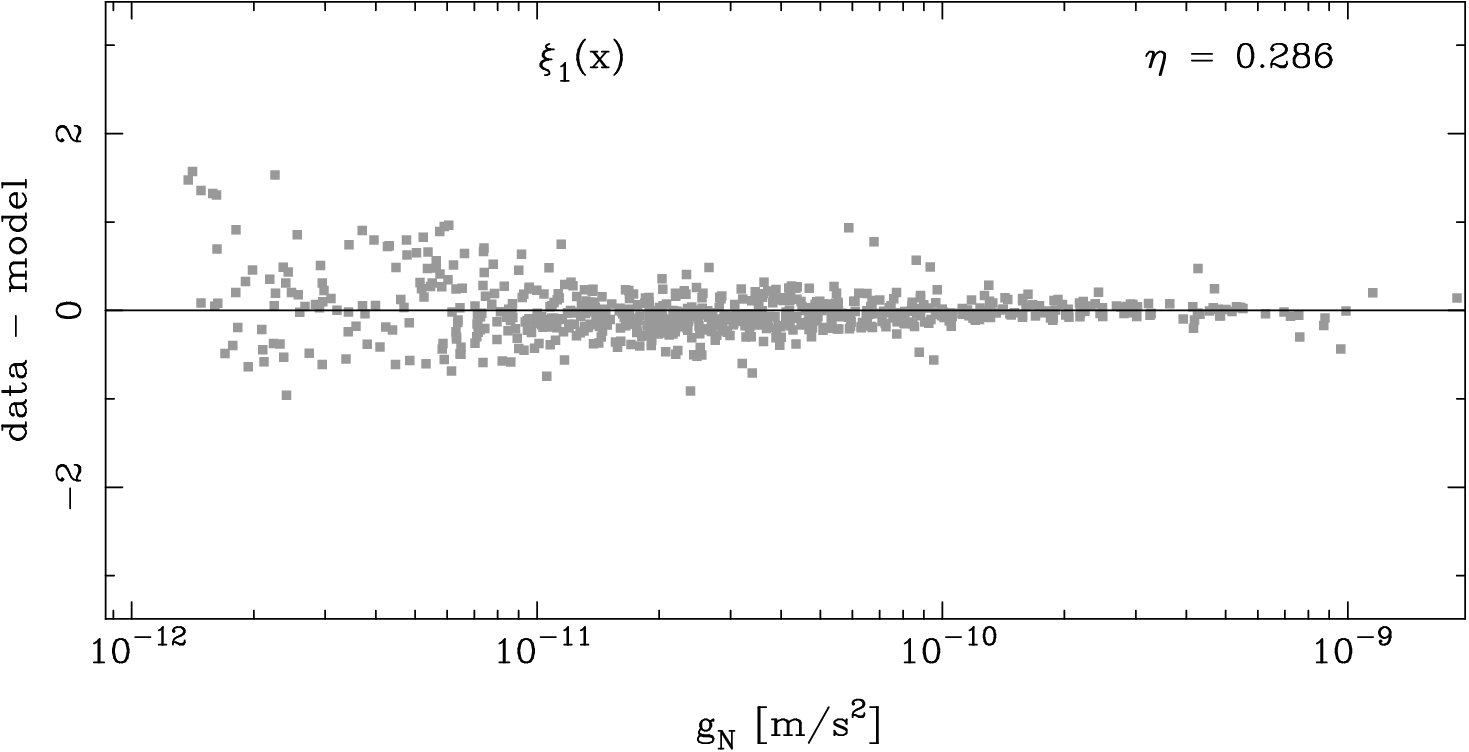} \\
\vspace{2mm}
\includegraphics[angle=-90,trim=0mm 0mm 15mm 0mm,clip,width=80mm]{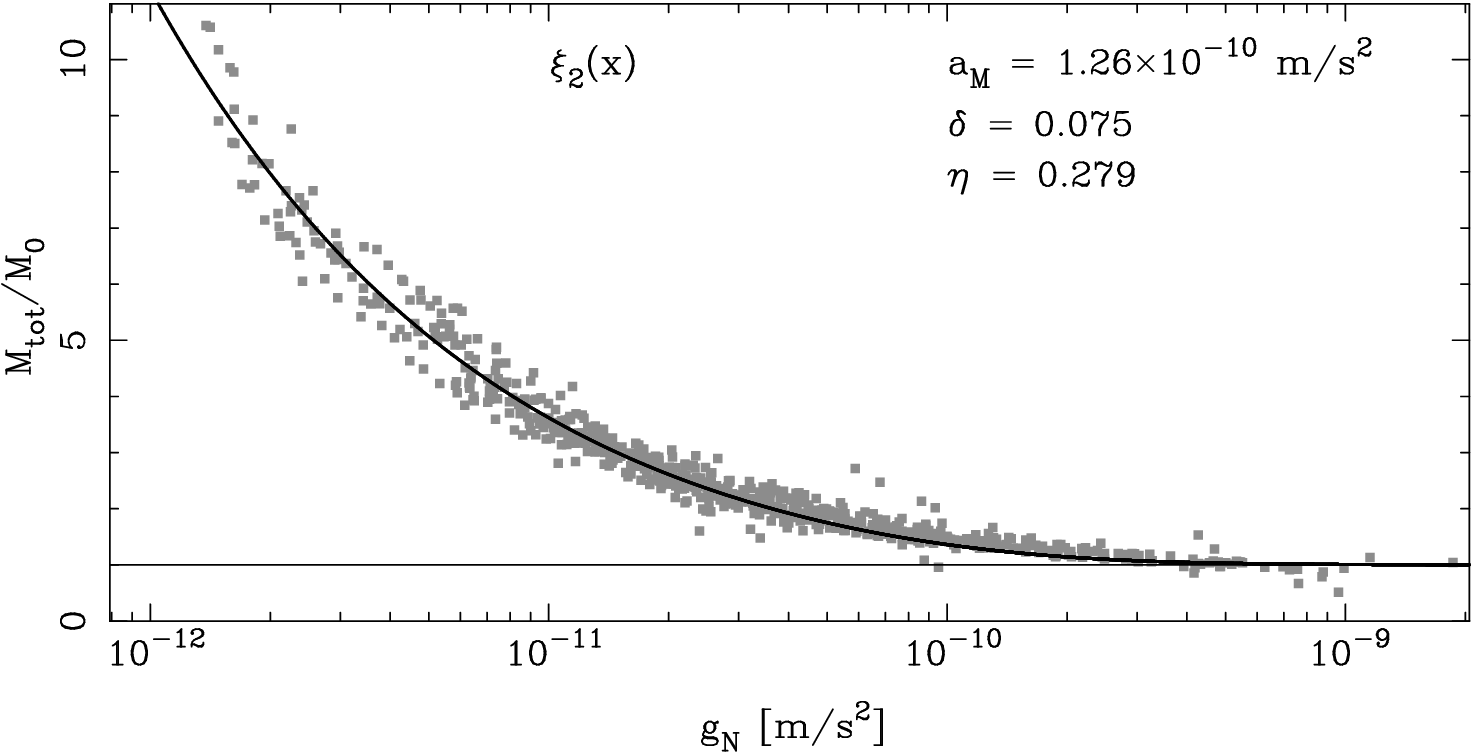}
\hspace{5mm}
\includegraphics[angle=-90,trim=0mm 0mm 15mm 0mm,clip,width=80mm]{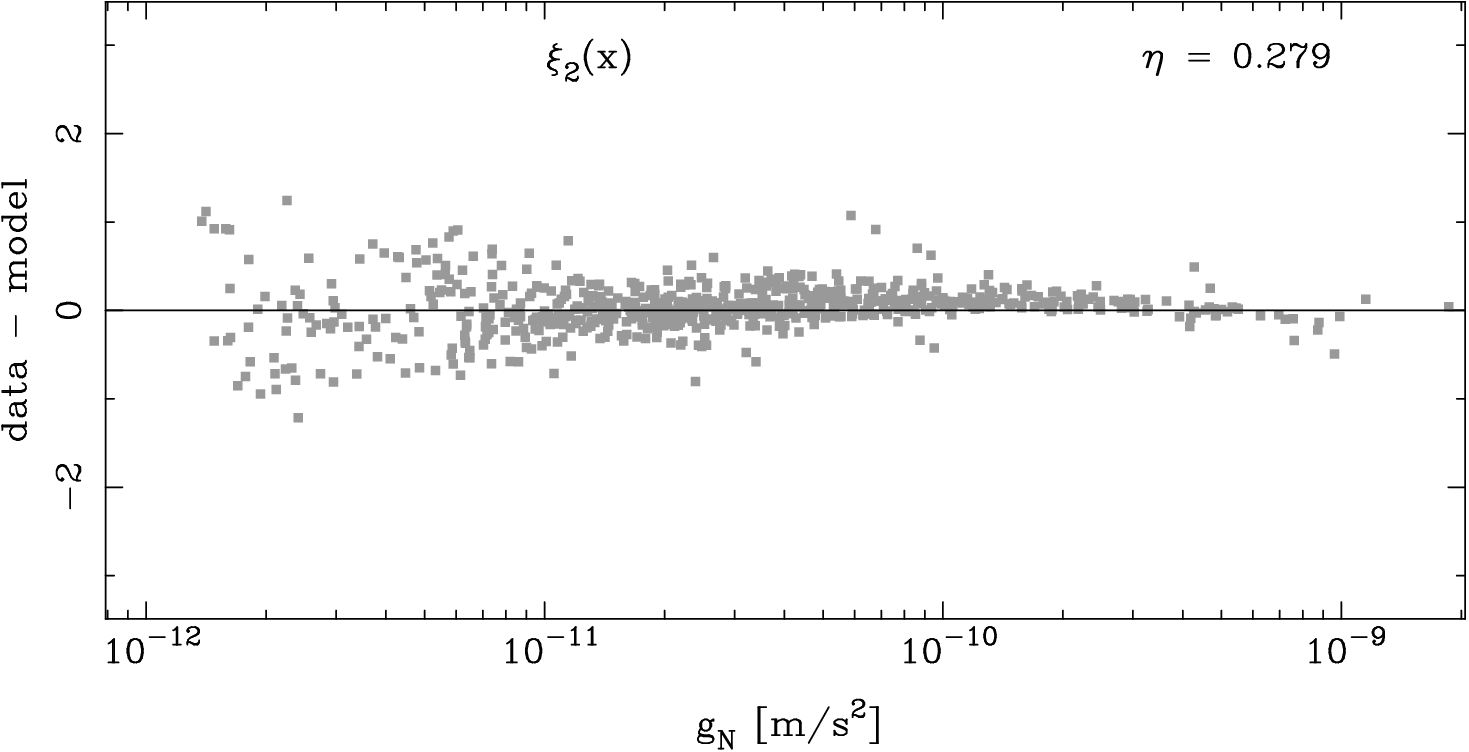} \\
\vspace{2mm}
\includegraphics[angle=-90,trim=0mm 0mm 15mm 0mm,clip,width=80mm]{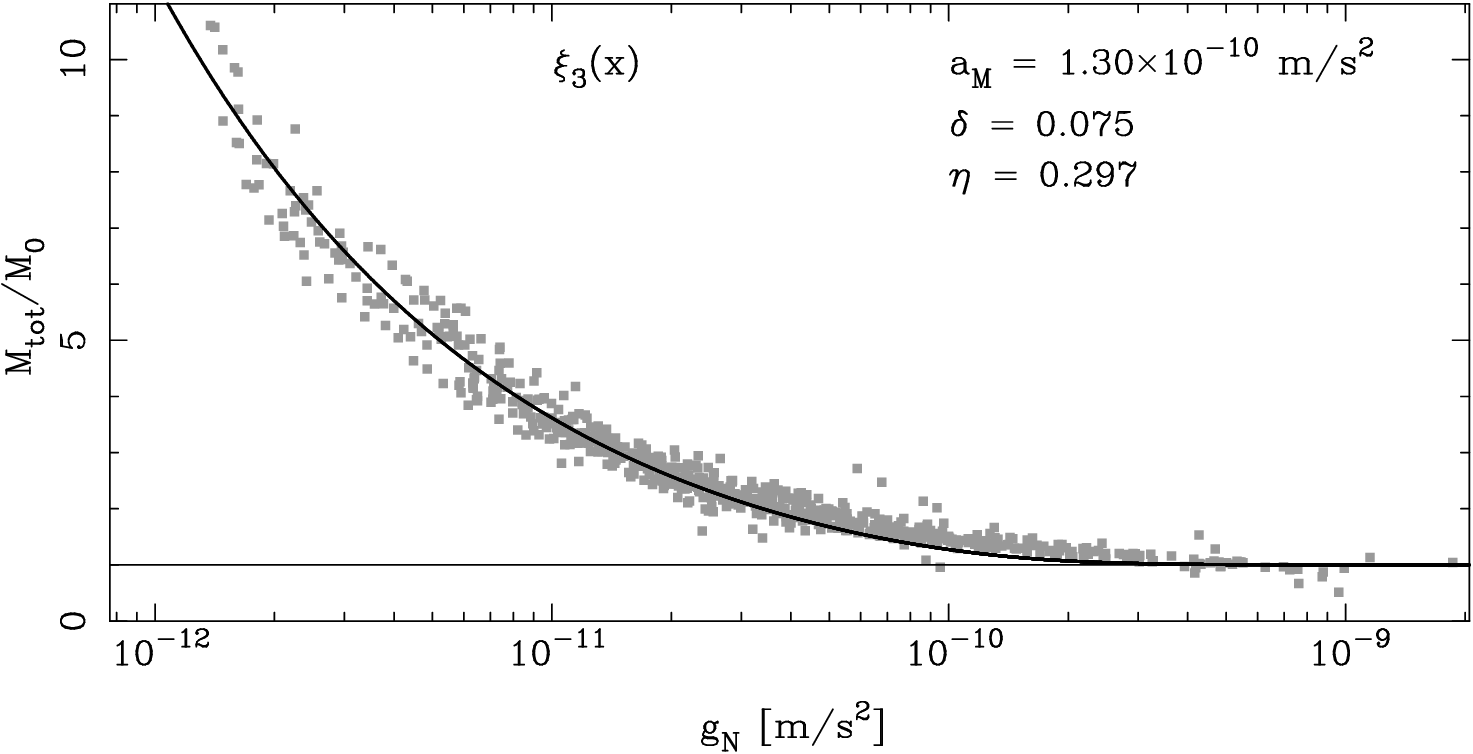}
\hspace{5mm}
\includegraphics[angle=-90,trim=0mm 0mm 15mm 0mm,clip,width=80mm]{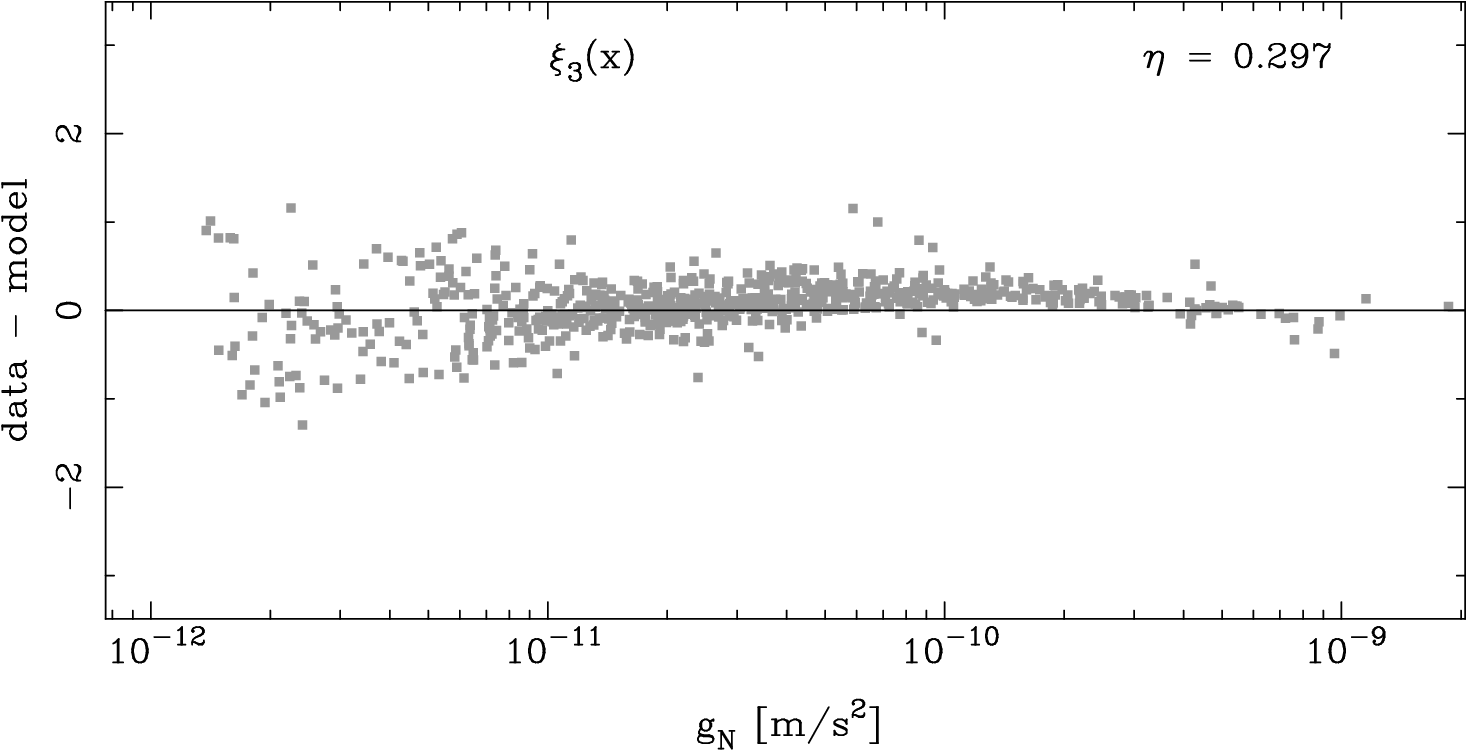} \\
\vspace{2mm}
\includegraphics[angle=-90,trim=0mm 0mm 0mm 0mm,clip,width=80mm]{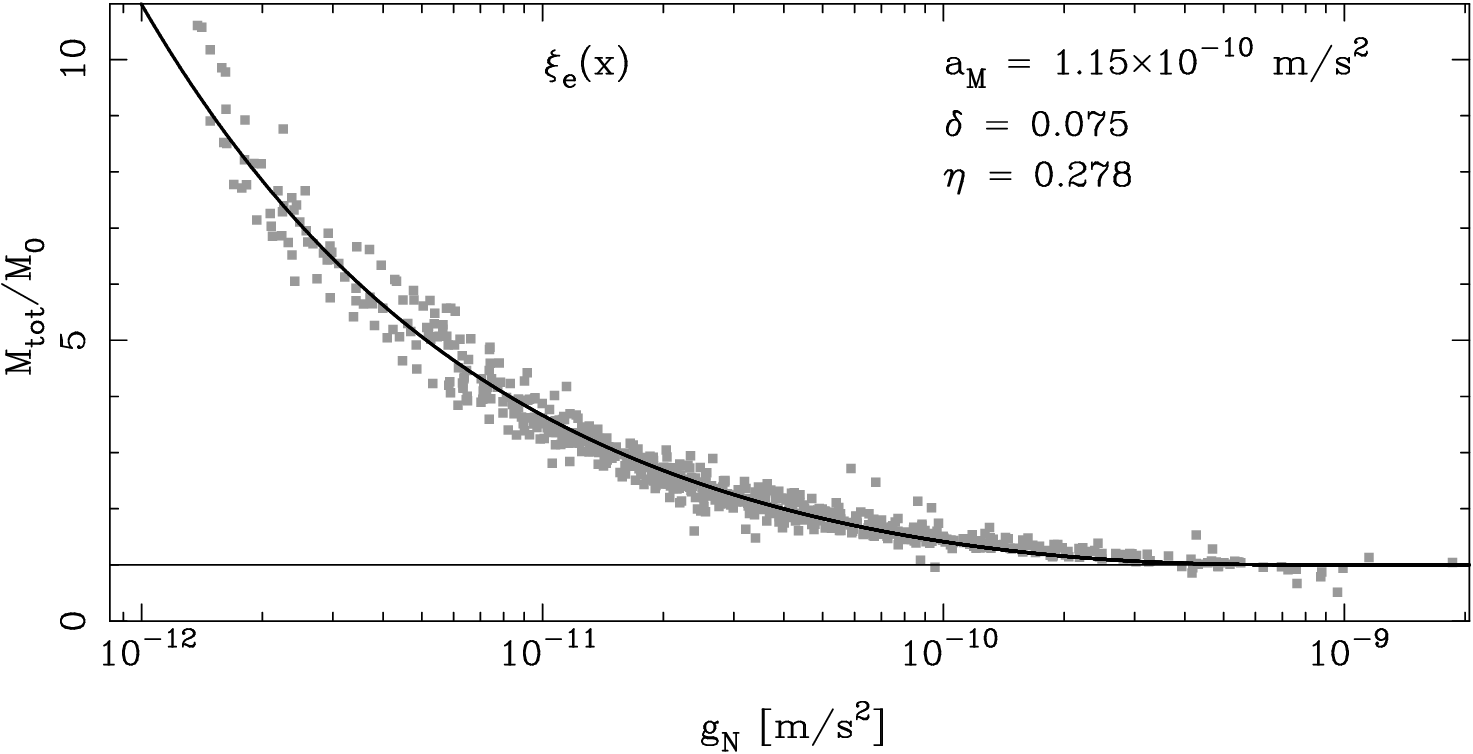}
\hspace{5mm}
\includegraphics[angle=-90,trim=0mm 0mm 0mm 0mm,clip,width=80mm]{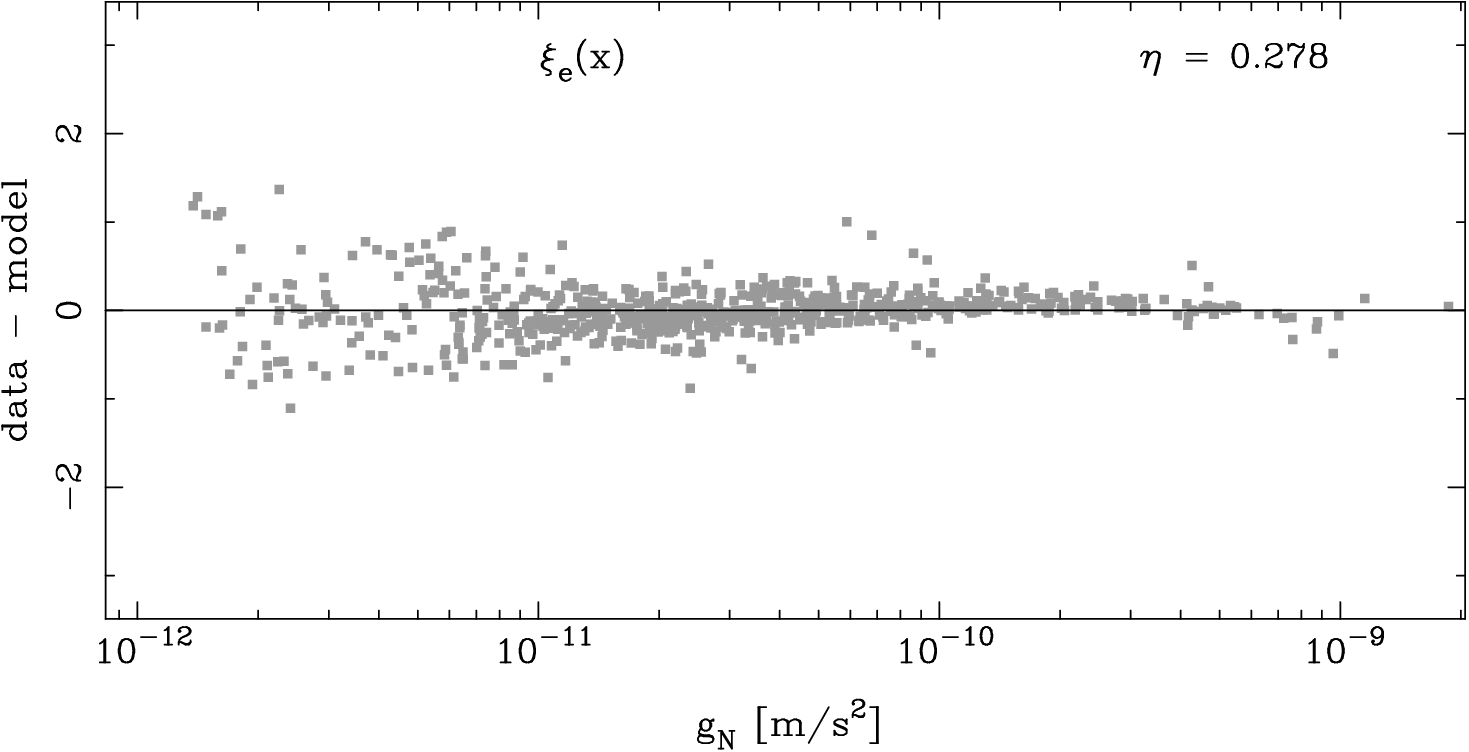} \\
\vspace{2mm}
\caption{Comparison of the theoretical MDA relations $\xii$, $\xiii$, $\xiiii$, and $\xie$ (from top to bottom). The \emph{left} column shows the best fits to the data, each diagram comprises the best-fit values of $\am$, $\delta$, and $\eta$. Grey points mark the MDA data (MD as function of $\gn$), black continuous curves indicate the best-fit models. The diagrams in the \emph{right} column show the corresponding residuals, i.e., the differences between data and best-fit models (in units of $\Mt/\Mo$). Please note the logarithmic--linear axis scales. \label{fig_res}}
\end{figure*}

\noindent
I illustrate the analysis of $\xii$ in Fig.~\ref{fig_mda}, showing both representations R1 and R2. A comparison of all four scaling laws is provided in Fig.~\ref{fig_res}, using R2 only for clarity. For each scaling law, I present a comparison of best-fit model and data (left column of Fig.~\ref{fig_res}) as well as the corresponding residual (right column of Fig.~\ref{fig_res}); I adopt the convention ``residual $=$ data $-$ model''. Each diagram provides the parameters $\am$, $\delta$, and $\eta$ of the corresponding best-fit model.

All scaling laws can be used to fit the data with more or less acceptable results. The post-fit r.m.s. residuals are

$\eta = 0.286$ for $\xii$,

$\eta = 0.279$ for $\xiii$,

$\eta = 0.297$ for $\xiiii$,

$\eta = 0.278$ for $\xie$,

\noindent
in units of $\Mt/\Mo$, meaning that the formal difference between the four models is quite moderate, with r.m.s. residuals differing by about 8\% in the most extreme case. The values found for Milgrom's constant are

$\am = 1.06\times10^{-10}$\,m\,s$^{-2}$ for $\xii$, 

$\am = 1.26\times10^{-10}$\,m\,s$^{-2}$ for $\xiii$,

$\am = 1.30\times10^{-10}$\,m\,s$^{-2}$ for $\xiiii$,

$\am = 1.15\times10^{-10}$\,m\,s$^{-2}$ for $\xie$,

\noindent
with offsets

$\delta = 0.225$ for $\xii$,

$\delta = 0.075$ for $\xiii$, $\xiiii$, and $\xie$,

\noindent
in units of $\Mt/\Mo$.

As pointed out in \S\,\ref{ssect_comparison}, the observational data are affected by systematic uncertainties that translate into (largely) systematic uncertainties in characteristic acceleration $\am$ and offset $\delta$; from the fitting procedure, I estimate these uncertainties to be about $0.05\times10^{-10}$\,m\,s$^{-2}$ and 0.05, respectively, for all scaling laws. I note that the bias affecting the data might be more complex than a simple offset $\delta$. However, I intentionally refrain from applying any more complex correction function because this comes with the risk of ``fitting the data to the model'' -- which is obviously unacceptable.

Even though the \emph{formal} r.m.s. fit residuals do not point toward or reject a certain scaling law in a straightforward manner, the \emph{systematic} residuals illustrated in the right column of Fig.~\ref{fig_res} do. Inspection of these residuals shows that $\xiiii$ does not provide an acceptable fit to the data and has to be rejected. The scaling laws $\xiii$ and $\xie$ provide better fits but show systematic deviations from the data at high accelerations $\gn\gtrsim10^{-10}$\,m\,s$^{-2}$, with $\xie$ performing better than $\xiii$; in both cases, the model curves drop toward zero too fast. Only the scaling law $\xii$ appears to be free of notable systematic deviations from the data. Interestingly, even though it provides the best fit, the r.m.s. residual of $\xii$ is about 4\% larger than the one for $\xie$; this is caused by $\xie$ providing a slightly better fit to the data at low accelerations $\gn\lesssim5\times10^{-12}$\,m\,s$^{-2}$ where the scatter is large.

From the analysis, I eventually conclude that $\xii$ provides a satisfactory description of the empirical MDA relation (see also Fig.~\ref{fig_mda}). It is free of notable systematic deviations from the data and provides the best fit -- though by a narrow margin -- of all four scaling laws tested. The value I find for Milgrom's constant, $\am=(1.06\pm0.05)\times10^{-10}$\,m\,s$^{-2}$, is in good agreement with independent estimates (e.g., \citealt{famaey2005,famaey2012}). Accordingly, my analysis is able to constrain Milgrom's constant $\am$ to within about 5\%.\footnote{Coincidentally, $\am=c\,H_0/2\pi$ within errors, with $c$ denoting the speed of light and $H_0\approx70$\,km\,s$^{-1}$\,Mpc$^{-1}$ being Hubble's constant (e.g., \citealt{riess2011,lee2012}). An approximate equality of $\am$ and $c\,H_0$ was already noted by \citet{milgrom1983a}.}

\section{DISCUSSION \label{sect_discuss}}

\noindent
The missing mass problem is commonly approached by postulating non-baryonic dark matter within the frame of $\Lambda$CDM cosmology. In recent years, it has become clear that this approach is incomplete. The assumption of dark matter distributed within and around galaxies is partially incompatible with observations of structure and kinematics of galaxies and groups of galaxies (see \citealt{kroupa2012} for a recent review). Standard cosmology seems unable to predict fundamental relations of galactic dynamics like the Faber-Jackson and Tully-Fisher relations \citep{faber1976,tully1977}. More generally, it has been found that galactic dynamics is intimately linked with a universal characteristic acceleration, Milgrom's constant $\am$, which shows up in the baryonic Tully-Fisher relation, the surface density--acceleration relation of galaxies, and the MDA relation \citep{famaey2012}. This can be understood in the frame of theories of modified gravity and/or inertia based on acceleration scales (e.g., \citealt{milgrom1983a,sanders2002,ferreira2009}).

The scaling law $\xii$, which can be derived from the ``graviton picture'' proposed in \citet{trippe2013a}, predicts a scaling of total dynamical mass $\Mt$ with centripetal acceleration $\ac$ according to Eq.~\ref{eq_mda}. Using $\vc^2=G\Mt/r$ and $\ac=\vc^2/r$ leads to

\begin{equation}
\label{eq_vc}
\vc^4 = G\,\Mo\,\am = const.
\end{equation}

\noindent
in the limit $\ac\ll\am$ as required by consistency with MOND; in the inverse case $\ac\gg\am$, the expression for the circular speed reduces to the usual Keplerian form. By construction, Eq.~\ref{eq_vc} is valid for \emph{rotationally supported} (dynamically cold) stellar systems. For \emph{pressure-supported} (dynamically hot) systems, one finds \citep{milgrom1984,milgrom1994}, again for $\ac\ll\am$,

\begin{equation}
\label{eq_sigma}
\sigma^4 = \frac{4}{9}\,G\,\Mo\,\am
\end{equation}

\noindent
with $\sigma$ being the three-dimensional stellar velocity dispersion (for isotropic velocity distributions and no rotation, $\sigma^2=3\sigma^2_z$, with $\sigma_z$ being the line-of-sight velocity dispersion). Given the motivation of this paper, it is important to note that Eq.~\ref{eq_sigma} does not depend on the functional form of the MDA relation but only on the limit expressed in Eq.~\ref{eq_vc}.

Equations~\ref{eq_vc} and \ref{eq_sigma} provide for the observed asymptotic flattening of galactic rotation curves and the Tully-Fisher/Faber-Jackson relations. When defining a luminous surface density $\Sigma\propto\Mo/r^2$ one finds $\ac\propto\Sigma^{1/2}$, in agreement with the empirical surface density--acceleration relation. However, the correct predictions of flat rotation curves, of the Tully-Fisher/Faber-Jackson relations, and of the surface density--acceleration relation do not depend on the functional form of the MDA relation but only on its asymptotic behavior in the limit of low accelerations $\ac\ll\am$. The new, and apparently unique, achievement of the ``graviton picture'' is a \emph{successful prediction of the functional form of the MDA relation and thus the transitional regime from Newtonian to modified dynamics.}

So far, I analyzed the scaling law $\xii$ with respect to the dynamics of disk galaxies; it is worth exploring for which dynamical systems and which ranges in mass and size it holds.

\noindent
{\bf Galaxy clusters:} Being the first dynamical systems for which a missing mass problem was noted \citep{zwicky1933}, clusters of galaxies have proven to be a challenge for dynamical studies ever since. Taking into account that most of the mass is not stored in stars but in the intra-cluster medium, observations find $(\Mt/\Mo)\approx8$ \citep{giodini2009}. The temperature $T$ of the intra-cluster medium is known to follow a relation $T^2\propto\sigma^4\propto\Mo$ as predicted by Eq.~\ref{eq_sigma}, i.e., a ``Faber--Jackson relation for galaxy clusters'' \citep{sanders1994}. The theoretical MD values still underestimate the observed ones by factors on the order of two (cf. \S\,6.6.4 of \citealt{famaey2012}). Given that present-day observations probably miss substantial amounts of intergalactic matter \citep{fukugita2004,shull2012}, meaning $\Mo$ is underestimated systematically, this discrepancy is unsurprising however.

\noindent
{\bf Galaxies:} As discussed in \S\,\ref{sect_results}, $\xii$ provides a good description of galactic dynamics over two orders of magnitude in $\ac$ and three orders of magnitude in $\gn$. This corresponds to a correct reproduction of rotation curves of disk galaxies on galactic radii between $\approx$1\,kpc and $\approx$80\,kpc (\citealt{sanders2002}; Fig.~23 of \citealt{famaey2012}; see also \citealt{famaey2005} for the specific case of the Milky Way). Equations~\ref{eq_vc} and \ref{eq_sigma} hold for at least seven orders of magnitude in mass, from dwarf galaxies to massive spirals (cf. Fig.~48 of \citealt{famaey2012}; see also \citealt{sanders2010} for the specific case of elliptical galaxies); including galaxy clusters, the range extends over ten orders of magnitude.

\noindent
{\bf Center of the Milky Way:} The Galactic center hosts a nuclear star cluster which includes the supermassive ($M_{\bullet}\approx4.3\times10^{6}\,M_{\odot}$; \citealt{gillessen2009}) black hole Sagittarius A* (Sgr~A*). The sphere of influence of Sgr~A* contains a total dynamical mass $\Mt\approx9\times10^6\,M_{\odot}$ within a radius $r\approx2$\,pc \citep{trippe2008,schoedel2009}. This implies that for the nuclear star cluster $(\Mt/\Mo)\approx1+3\times10^{-4}$; the deviation from unity predicted by $\xii$ is about three orders of magnitude below the sensitivity of current dynamical studies (e.g., \citealt{trippe2008}). For the other three scaling laws, the deviations from unity are yet smaller by several orders of magnitude.

\noindent
{\bf Globular clusters:} Located in the outer Galactic halo, several isolated globular star clusters have received attention recently as test benches for theories of gravity. Various studies (e.g., \citealt{scarpa2011,hernandez2013}) find that the stellar velocity dispersions of globular clusters approach constant values $\sigma\propto\Mo^{1/4}$ asymptotically at large radii (and thus low accelerations), in agreement with Eq.~\ref{eq_sigma} and in disagreement with standard Newtonian dynamics. More detailed comparisons between theoretical and observed cluster masses and velocity dispersions have been inconclusive as yet (cf., e.g., \citealt{frank2012} for a recent study of the cluster Palomar 4): as pointed out by \citet{sanders2012}, the theoretically expected cluster kinematics strongly depends on the initial assumptions made for the phase-space distribution of the cluster stars, especially if the cluster is approximated as an isothermal sphere or not.

\noindent
{\bf Binary stars:} For stellar masses $M_{\star}\equiv\Mo\approx1\,M_{\odot}$, the centripetal acceleration experienced by a test particle orbiting the star at radial distance $r$ reaches the regime $\ac\lesssim\am$ for $r\gtrsim7000$\,AU. Accordingly, wide binary star systems can be used to probe the validity of Eq.~\ref{eq_vc}. Based on a sample of 417 wide binaries, \citet{hernandez2012} find that the circular velocities of the companion stars are independent of their separations, in agreement with Eq.~\ref{eq_vc} and in disagreement with Kepler's third law.

\noindent
{\bf Solar system:} The solar system arguably provides the strongest constraints on any theory of gravity. All MDA scaling laws discussed in this work imply an increase of the effective mass of the sun experienced by the solar planets. The observational limits on $(\Mt/\Mo)-1$ experienced by Jupiter, Uranus, and Neptune, in units of $10^{-7}$, are 2, 18, and 35, respectively ($3\sigma$ confidence levels; \citealt{anderson1995}). The values predicted by the scaling laws $\xiii$, $\xiiii$, and $\xie$ are lower than the observational limits by many orders of magnitude  (using orbital data from \citealt{tholen2000}). However, the values predicted by $\xii$ are 5, 65, and 159, respectively -- exceeding the observational limits by factors of three to five.

Regarding the combined evidence, I can draw several conclusions. First, none of the scaling laws I examine is truly universal in the sense of being valid on \emph{all} scales, from the solar system to clusters of galaxies. Second, the scaling law $\xii$ provides the best fit to the empirical MDA relation; including the limiting cases expressed in Eqs. \ref{eq_vc} and \ref{eq_sigma}, it provides a good and simple description of galactic dynamics on all scales. Third, the apparent inconsistency of $\xii$ with solar system kinematics -- i.e., for a small dynamical system with $\ac\gg\am$ -- clearly indicates that $\xii$ is only an approximation of a more sophisticated scaling law.

The interesting feature of $\xii$ is that it is physically motivated; it can be derived from a ``graviton picture'' of gravity. Even though the idea of gravity being mediated by massive particles -- ``massive gravity'' -- has not been applied to galactic dynamics yet, it is far from new. As already noted by \citet{fierz1939}, gravitation can be described as being mediated by a virtual particle with spin 2, a \emph{graviton}, with a mass $\mg\geq0$. Since then, massive gravity has been an area of active research theoretically as well as experimentally (see, e.g., \citealt{goldhaber2010,hinterbichler2012} for reviews). Multiple experiments have been able to constrain the graviton mass to $\mg<10^{-65}\,{\rm kg}\approx10^{-29}$\,eV (lowest model-independent limit; \citealt{goldhaber2010}). The possibility of non-zero graviton masses has found application in cosmology in view of the \emph{dark energy problem}: Heisenberg's uncertainty relation demands that virtual particles with non-zero mass have limited life times. Accordingly, $\mg>0$ implies an exponential decay of gravity on cosmological scales, corresponding to an apparent accelerated expansion of the universe (e.g., \citealt{hinterbichler2012}). Recently, \citet{cardone2012} reported that massive gravity is indeed consistent with cosmological observations.

From the discussion above follows that the assumption of massive gravity is, a priori, a possibility to be considered in the context of galactic dynamics. Indeed, the ``graviton picture'' is sufficient to describe stellar dynamics on all scales -- however, this does not imply that it is necessary or is a unique approach to the MDA relation. Evidently, the simple toy model I use in \S\,\ref{ssect_theomda}, which is essentially based on classical mechanics, can only be a first-order approximation to a more sophisticated, relativistic theory of gravity. Indeed, relativistic formulations of MONDian gravitation, notably Bekenstein's \emph{tensor--vector--scalar theory} (TeVeS), have been known for almost a decade -- however with the remarkable limitation of not providing the MDA relation a priori but assuming various scaling laws ad hoc \citep{bekenstein2004,bekenstein2006}. Eventually, I end up with the following careful conclusion: On galactic scales, gravity behaves as if it was mediated by massive particles -- gravitons.

\section{CONCLUSIONS}

\noindent
I compare the empirical mass discrepancy--acceleration relation of disk galaxies first reported by \citet{mcgaugh2004} with four theoretical scaling laws. One of these scaling laws, $\xii$, can be derived from a ``graviton picture'' of gravity \citep{trippe2013a,trippe2013b}, the other three are empirically motivated. I arrive at the following principal conclusions:

\begin{enumerate}

\item  The scaling law $\xii$ provides the best fit to the MDA data. It comprises one free parameter, Milgrom's constant, which I find to be $\am=1.06\times10^{-10}$\,m\,s$^{-2}$ within a (largely systematic) relative uncertainty of 5\%.

\item  In addition to the successful prediction of the MDA relation, the ``graviton picture'' is consistent with stellar dynamics on all scales, from Galactic binary stars to clusters of galaxies, thus covering stellar systems spanning eight orders of magnitude in size and 14 orders of magnitude in mass.

\item  The scaling law $\xii$ is probably inconsistent with solar system kinematics, indicating that it is only an approximation of a more sophisticated law of gravity.

\end{enumerate}

Regarding the combined evidence, the ``graviton picture'' provides a good and simple description of galactic dynamics on all scales despite its toy-model character. This suggests that, on galactic scales, gravity behaves as if it was mediated by massive particles -- gravitons.

\acknowledgments{\noindent I am grateful to {\small\sc Stacy S. McGaugh} (Case Western Reserve Univ.) for making available the MDA data set and for valuable discussion, and to {\small\sc Junghwan Oh}, {\small\sc Taeseok Lee}, {\small\sc Jae-Young Kim}, and {\small\sc Jong-Ho Park} (all at SNU) for technical support. This work made use of the software package {\sc dpuser} developed and maintained by {\small\sc Thomas Ott} at MPE Garching.\footnote{{\tt http://www.mpe.mpg.de/$\sim$ott/dpuser/dpuser.html}} I acknowledge financial support from the Korean National Research Foundation (NRF) via Basic Research Grant 2012R1A1A2041387. Last but not least, I am grateful to an anonymous reviewer for valuable comments.}


\begin{thebibliography}{}
\setlength{\itemsep}{-0.1mm}
\small

\bibitem[Anderson \etal(1995)]{anderson1995} Anderson, J.D., et al. 1995, Improved Bounds on Nonluminous Matter in Solar Orbit, ApJ, 448, 885
\bibitem[Bahcall \etal(1999)]{bahcall1999} Bahcall, N.A., Ostriker, J.P., Perlmutter, S. \& Steinhardt, P.J. 1999, The Cosmic Triangle: Revealing the State of the Universe, Science, 284, 1481
\bibitem[Bekenstein \& Milgrom(1984)]{bekenstein1984} Bekenstein, J.D. \& Milgrom, M. 1984, Does the Missing Mass Problem Signal the Breakdown of Newtonian Gravity?, ApJ, 286, 7
\bibitem[Bekenstein(2004)]{bekenstein2004} Bekenstein, J.D. 2004, Relativistic Gravitation Theory for the Modified Newtonian Dynamics Paradigm, Phys. Rev. D, 70, 083509-1
\bibitem[Bekenstein(2006)]{bekenstein2006} Bekenstein, J.D. 2006, The Modified Newtonian Dynamics---MOND and Its Implications for New Physics, Contemp. Phys., 47, 387
\bibitem[Binney \& Tremaine(1987)]{binney1987} Binney, J. \& Tremaine, S. 1987, Galactic Dynamics, Princeton Univ. Press, Princeton
\bibitem[Cardone, Radicella \& Parisi(2012)]{cardone2012} Cardone, V.F., Radicella, N. \& Parisi, L. 2012, Constraining Massive Gravity with Recent Cosmological Data, Phys. Rev. D, 85, 124005
\bibitem[Einasto \etal(1974)]{einasto1974} Einasto, J., Kaasik, A. \& Saar, E. 1974, Dynamic Evidence on Massive Coronas of Galaxies, Nature, 250, 309
\bibitem[Faber \& Jackson(1976)]{faber1976} Faber, S.M. \& Jackson, R.E. 1976, Velocity Dispersions and Mass-to-Light Ratios for Elliptical Galaxies, ApJ, 204, 668
\bibitem[Famaey \& Binney(2005)]{famaey2005} Famaey, B. \& Binney, J. 2005, Modified Newtonian Dynamics in the Milky Way, MNRAS, 363, 603
\bibitem[Famaey \& McGaugh(2012)]{famaey2012} Famaey, B. \& McGaugh, S.S. 2012, Modified Newtonian Dynamics (MOND): Observational Phenomenology and Relativistic Extensions, Living Rev. Relativ., 15, 10
\bibitem[Ferreira \& Starkman(2009)]{ferreira2009} Ferreira, P.G. \& Starkman, G.D. 2009, Einstein's Theory of Gravity and the Problem of Missing Mass, Science, 326, 812
\bibitem[Fierz \& Pauli(1939)]{fierz1939} Fierz, M. \& Pauli, W. 1939, On Relativistic Wave Equations for Particles of Arbitrary Spin in an Electromagnetic Field, Proc. R. Soc. London A, 173, 211
\bibitem[Frank \etal(2012)]{frank2012} Frank, M.J., et al. 2012, The Velocity Dispersion and Mass Function of the Outer Halo Globular Cluster Palomar 4, MNRAS, 423, 2917
\bibitem[Fukugita \& Peebles(2004)]{fukugita2004} Fukugita, M. \& Peebles, P.J.E. 2004, The Cosmic Energy Inventory, ApJ, 616, 643
\bibitem[Gallagher \& Hudson(1976)]{gallagher1976} Gallagher, J.S. \& Hudson, H.S. 1976, Surface Photometry of the Spiral Galaxy IC\,2233 and the Existence of Massive Halos, ApJ, 209, 389
\bibitem[Gillessen \etal(2009)]{gillessen2009} Gillessen, S., Eisenhauer, F., Trippe, S. et al. 2009, Monitoring Stellar Orbits around the Massive Black Hole in the Galactic Center, ApJ, 692, 1075
\bibitem[Giodini \etal(2009)]{giodini2009} Giodini, S., et al. 2009, Stellar and Total Baryon Mass Fractions in Groups and Clusters Since Redshift 1, ApJ, 703, 982
\bibitem[Goldhaber \& Nieto(2010)]{goldhaber2010} Goldhaber, A.S. \& Nieto, M.M. 2010, Photon and Graviton Mass Limits, Rev. Mod. Phys., 82, 939
\bibitem[Griffith(2008)]{griffith2008} Griffith, D. 2008, Introduction to Elementary Particles, Wiley-VCH, Weinheim
\bibitem[Hernandez \etal(2012)]{hernandez2012} Hernandez, X., Jim\'enez, M.A. \& Allen, C. 2012, Wide Binaries as a Critical Test of Classical Gravity, Eur. Phys. J. C, 72, 1884
\bibitem[Hernandez \etal(2013)]{hernandez2013} Hernandez, X., Jim\'enez, M.A. \& Allen, C. 2013, Flattened Velocity Dispersion Profiles in Globular Clusters: Newtonian Tides or Modified Gravity?, MNRAS, 428, 3196
\bibitem[Hinterbichler(2012)]{hinterbichler2012} Hinterbichler, K. 2012, Theoretical Aspects of Massive Gravity, Rev. Mod. Phys., 84, 671
\bibitem[Kroupa(2012)]{kroupa2012} Kroupa, P. 2012, The Dark Matter Crisis: Falsification of the Current Standard Model of Cosmology, PASA, 29, 395
\bibitem[Lee \& Jang(2012)]{lee2012} Lee, M.G. \& Jang, I.S. 2012, The Distance to M\,101 Hosting Type Ia Supernova 2011fe Based on the Tip of the Red Giant Branch, ApJL, 760, L14
\bibitem[McGaugh(2004)]{mcgaugh2004} McGaugh, S.S. 2004, The Mass Discrepancy--Acceleration Relation: Disk Mass and the Dark Matter Distribution, ApJ, 609, 652
\bibitem[Milgrom(1983a)]{milgrom1983a} Milgrom, M. 1983, A Modification of the Newtonian Dynamics as a Possible Alternative to the Hidden Mass Hypothesis, ApJ, 270, 365
\bibitem[Milgrom(1983b)]{milgrom1983b} Milgrom, M. 1983, A Modification of the Newtonian Dynamics: Implications for Galaxies, ApJ, 270, 371
\bibitem[Milgrom(1983c)]{milgrom1983c} Milgrom, M. 1983, A Modification of the Newtonian Dynamics: Implications for Galaxy Systems, ApJ, 270, 384
\bibitem[Milgrom(1984)]{milgrom1984} Milgrom, M. 1984, Isothermal Spheres in the Modified Dynamics, ApJ, 287, 571
\bibitem[Milgrom(1994)]{milgrom1994} Milgrom, M. 1994, Modified Dynamics Predictions Agree with Observations of the H\,{\sc i} Kinematics in Faint Dwarf Galaxies Contrary to the Conclusions of Lo, Sargent, and Young, ApJ, 429, 540
\bibitem[Ostriker \& Peebles(1973)]{ostriker1973} Ostriker, J.P \& Peebles, P.J.E. 1973, A Numerical Study of the Stability of Flattened Galaxies: Or, Can Cold Gas Survive?, ApJ, 186, 467
\bibitem[Riess \etal(2011)]{riess2011} Riess, A.G., et al. 2011, A 3\% Solution: Determination of the Hubble Constant with the Hubble Space Telescope and Wide Field Camera 3, ApJ, 730, 119
\bibitem[Rubin \etal(1980)]{rubin1980} Rubin, U.C., Ford, W.K. Jr. \& Thonnard, N. 1980, Rotational Properties of 21 Sc Galaxies with a Large Range of Luminosities and Radii, From NGC\,4605 (R = 4\,kpc) to UGC\,2885 (R = 122\,kpc), ApJ, 238, 471
\bibitem[Sanders(1990)]{sanders1990} Sanders, R.H. 1990, Mass Discrepancies in Galaxies: Dark Matter and Alternatives, A\&AR, 2, 1
\bibitem[Sanders(1994)]{sanders1994} Sanders, R.H. 1994, A Faber--Jackson Relation for Clusters of Galaxies: Implications for Modified Dynamics, A\&A, 284, L31
\bibitem[Sanders(2010)]{sanders2010} Sanders, R.H. 2010, The Universal Faber--Jackson Relation, MNRAS, 407, 1128
\bibitem[Sanders(2012)]{sanders2012} Sanders, R.H. 2012, NGC\,2419 Does Not Challenge Modified Newtonian Dynamics, MNRAS, 419, L6
\bibitem[Sanders \& McGaugh(2002)]{sanders2002} Sanders, R.H. \& McGaugh, S.S. 2002, Modified Newtonian Dynamics as an Alternative to Dark Matter, ARA\&A, 40, 263
\bibitem[Scarpa \etal(2011)]{scarpa2011} Scarpa, R., et al. 2011, Testing Newtonian Gravity with Distant Globular Clusters: NGC\,1851 and NGC\,1904, A\&A, 525, A148
\bibitem[Sch\"odel, Merritt \& Eckart(2009)]{schoedel2009} Sch\"odel, R., Merritt, D. \& Eckart, A. 2009, The Nuclear Star Cluster of the Milky Way: Proper Motions and Mass, A\&A, 502, 91
\bibitem[Shull, Smith \& Danforth(2012)]{shull2012} Shull, J.M., Smith, B.D. \& Danforth, C.W. 2012, The Baryon Census in Multiphase Intergalactic Medium: 30\% of the Baryons May Still Be Missing, ApJ, 759, 23
\bibitem[Tholen, Tejfel \& Cox(2000)]{tholen2000} Tholen, D.J., Tejfel, V.G. \& Cox, A.N. 2000, in: Cox, A.N. (ed.), Allen's Astrophysical Quantities, 4th edn., Springer, New York, 293
\bibitem[Trippe \etal(2008)]{trippe2008} Trippe, S., et al. 2008, Kinematics of the Old Stellar Population at the Galactic Centre, A\&A, 492, 419
\bibitem[Trippe(2013a)]{trippe2013a} Trippe, S. 2013, A Simplified Treatment of Gravitational Interaction on Galactic Scales, JKAS, 46, 41
\bibitem[Trippe(2013b)]{trippe2013b} Trippe, S. 2013, A Derivation of Modified Newtonian Dynamics, JKAS, 46, 93
\bibitem[Tully \& Fisher(1977)]{tully1977} Tully, R.B. \& Fisher, J.R. 1977, A New Method of Determining Distances to Galaxies, A\&A, 54, 661
\bibitem[Zwicky(1933)]{zwicky1933} Zwicky, F. 1933, Die Rotverschiebung von extragalaktischen Nebeln, Helv. Phys. Acta, 6, 110 

\end{thebibliography}
\end{document}